\documentclass[aps,amsmath,amsfonts,twocolumn]{revtex4-1}%
\usepackage{mathtools}
\usepackage{verbatim}
\usepackage{hyperref}
\usepackage{epsfig,color,amsfonts,amsmath,amsthm,comment,natbib}
\usepackage{color}
\usepackage{graphicx}
\usepackage{braket}
\usepackage{amsthm}
\usepackage{amssymb}
\usepackage{amsmath}
\usepackage{ulem}
\usepackage{tikz}
\usepackage{subcaption}

\newcommand{\integers}{\mathbb{Z}}

\newcommand{\IK}{\color{red}}

\newcommand{\MW}{\color{blue}}

\newcommand{\old}{\color{black}}

\begin{document}

\title{Arrested development and fragmentation in Strongly-Interacting Floquet Systems}

\author{Matthew Wampler$^{1}$ and Israel Klich$^{1}$}
\affiliation{$^1$Department of Physics, University of Virginia, Charlottesville, Virginia 22903, USA
}

\begin{abstract}
We explore how interactions can facilitate classical like dynamics in models with sequentially activated hopping. Specifically, we add local and short range interaction terms to the Hamiltonian, and ask for conditions ensuring the evolution acts as a permutation on initial local number Fock states. We show that at certain values of hopping and interactions, determined by a set of Diophantine equations, such evolution can be realized. When only a subset of the Diophantine equations is satisfied the Hilbert space can be fragmented into frozen states, states obeying cellular automata like evolution and subspaces where evolution mixes Fock states and is associated with eigenstates exhibiting high entanglement entropy and level repulsion. 
\end{abstract}


\maketitle
\section{Introduction}
As experimental tools have progressed (e.g. \cite{Bloch2008Ultracold,Blatt2012TrappedIon}), the microscopic control of quantum systems has become increasingly accessible.  These advancements, along with a correlated increase in theoretical interest, have led to the discovery of many new and surprising phenomena that emerge when periodic driving, interactions, and their interplay are considered.  

For example, periodically driven systems can be used
to stabilize otherwise unusual behavior. A recent
important example is topological Floquet insulators \cite{Kitagawa2010FTI,lindner2011Floquet,Rudner2020FTI}, 
where novel topological features of the band structure may emerge
due to inherent periodicity of the non-interacting quasi-
energy spectrum.  Furthermore, it was shown in \cite{titum2016anomalous} that, by combining spatial disorder with a topological Floquet insulator model introduced by Rudner-Lindner-Berg-Levin (RLBL) \cite{rudner2013anomalous}, a new topological phase may be realized called the anomalous Floquet-Anderson insulator (AFAI).   %The AFAI is anomalous in the sense that it supports properties impossible to realize in systems with static, local Hamiltonians.  Namely, it supports chiral edge modes and a fully localized bulk simultaneously.  The anomalous, topological edge behavior has been realized in photonic, acoustic, and ultracold atomic systems \cite{Peng2016AFAIExp,Maczewsky2017AFAIExp,Mukherjee2017AFAIExp,Wintersperger2020AFAIExp}, while a realization of the full AFAI phase was recently proposed for solid state systems \cite{Zhang2022AFAIExp}.  
Discrete time crystals \cite{Sacha2020DTCBook,Else2020DTCRev,Khemani2016DTC,Else2016DTC} are another important example of behavior that may occur in periodically driven, but not static \cite{Watanabe2015noTC}, systems.  Namely, a time crystal is a system where time-translation symmetry is spontaneously broken (in analogy to spatial translation symmetry spontaneously breaking to form ordinary crystals).

Combining periodic driving with interactions, however, can often be problematic as generic, clean, interacting Floquet system are
expected to indefinitely absorb energy from their drive
and thus quickly converge to a featureless infinite temperature
state \cite{Lazarides2014Therm,Dalessio2014Therm,Ponte2015Therm}. This problem may be side-stepped by
considering Many-Body Localization (MBL) \cite{Abanin2019MBLRev,DAlessio2013MBL,Ponte2015MBL,Ponte2015MBLPRL,Lazarides2015MBL,Khemani2016MBL,Agarwala2017MBL}, in which strong
disorder is utilized to help stave off thermalization,
by considering the effective evolution of pre-thermal
states \cite{Bukov2015pretherm,Kuwahara2016pretherm,Else2017pretherm,Abanin2017pretherm,Zeng2017pretherm,Machado2019pretherm} that, in the best cases, take exponentially
long to thermalize, or by connecting the system to a
bath to facilitate cooling and arrive at interesting, non-
equilibrium steady-states \cite{Dehghani2014Dissipative,Iadecola2015Bath,Iadecola2015Bath2,Seetharam2015Bath}.

Yet another route for realizing non-trivial dynamics despite the expected runaway heating from interacting, Floquet drives is to consider systems where the ergodicity is weakly broken, i.e. where there are subspaces (whose size scales only polynomially in the system size) of the Hilbert space that do not thermalize despite the fact that the rest of the Hilbert space does.  These non-thermal states are called quantum many-body scars \cite{Turner2018Scars,Ho2019Scars,Moudgalya2022Scars} and have been shown to support many interesting phenomena including, for example, discrete time crystals \cite{Yarloo2020ScarDTC}.  Furthermore, in constrained systems, the full Hilbert space may fragment into subspaces where some of the subspaces thermalize while others do not \cite{Sala2020HFrag,Moudgalya2022Scars}.  When the fraction of non-thermal states are a set of measure zero in the thermodynamic limit, the system is an example of quantum many-body scarring.  However, in other cases, the non-thermal subspaces form a finite fraction of the full Hilbert space and therefore correspond to a distinct form of ergodicity breaking.        

In addition to leading to heating, interactions are also often responsible for our inability to efficiently study or describe many body quantum states in both Floquet and static Hamiltonian systems. However, there are situations when interactions play the opposite role in creating specialized states of particular simplicity or utility. For example, systems with interactions can exhibit  counter-intuitive bound states due to coherent blocking of evolution. A nice class of such systems are the edge-locked few particle systems studied in \cite{haque:060401,haque2010self}. 

In this work, we consider Floquet drives where hopping between neighboring pairs of sites are sequentially activated.  The theoretical and experimental tractability of such models have made them a popular workhorse for fleshing out a broad range of the exciting properties of periodically driven systems (e.g. \cite{rudner2013anomalous,Kumar2018evenodd,Ljubotina2019evenodd,Piroli2020evenodd,Lu2022EvenOddPRL}).  We find that, when interactions are added to such systems, there exist special values of interaction strength and driving frequency where the dynamics becomes exactly solvable.  Furthermore, the complete set of these special parameter values may be determined via emergent Diophantine equations \cite{Cohen2007Dioph}.  At other parameter values, the Hilbert space is fragmented.  Initial states contained within some, thermal, subspaces will ergodically explore the subspace (though not the entire Hilbert space), while other initial states contained within other, non-thermal, subspaces will evolve according to a classical cellular automation (CA) \cite{Wolfram1983CA}, i.e. the system evolves in discrete time steps where after each step the occupancy of any given site is updated deterministically based on a small set of rules determined by the occupancy of neighboring sites.

As examples, we consider RLBL(-like) models with added nearest neighbor (NN) or Hubbard interactions as well as an even-odd Floquet drive in one dimension with NN interactions (more detailed descriptions of these models given below).  We note that some work has been done in the first two cases \cite{Nathan2019AFI,Nathan2021AFI} where it was argued that novel, MBL anomalous Floquet insulating phases emmerged when a disorder potential was added.  We will discuss how our focus on special parameter values leads to new insights into these models and how it suggests a possible route towards other exciting phenomena such as the support of discrete time crystals within fragments of the Hilbert space.    
%In this work, we provide further investigations into the properties of the RLBL model with NN and Hubbard interactions (henceforth referred to as NN-RLBL and Hubbard-RLBL for brevity).  We find that, at certain values of driving frequency and interaction strength, the evolution becomes exactly solvable.  Specifically, the system evolves stroboscopically as a cellular automation \cite{Wolfram1983CA}, i.e. the system evolves in discrete time steps where after each step the occupancy of any given site is updated deterministically based on a small set of rules determined by the occupancy of neighboring sites.  The methods used to find classically evolving regions in parameter space of the Hubbard- and NN-RLBL models, can be leveraged to find classically evolving regions for other classes of interacting, periodically driven models, and in particular, it is independent of the dimension of the system.   

\IK
\old
\section{Conditions for evolution by Fock state permutations} 

In this section, we examine conditions for deterministic evolution of Fock states into Fock states in fermion models. Here we consider real space Fock states, which have a well defined fermion occupation on each lattice site (We will also refer to such states as fermion product states). We consider models where hopping between non-overlapping selected pairs of sites is sequentially activated. Two models of this type, discussed in detail below, deal with Hubbard and nearest neighbour interactions. The approach can be naturally extended to deal with more general interactions in sequentially applied evolution models.

\subsection{Example 1: Hubbard-RLBL}
\label{Section: Hubbard RLBL}
\begin{figure}[h]
    \centering
    \includegraphics[width=0.4\textwidth]{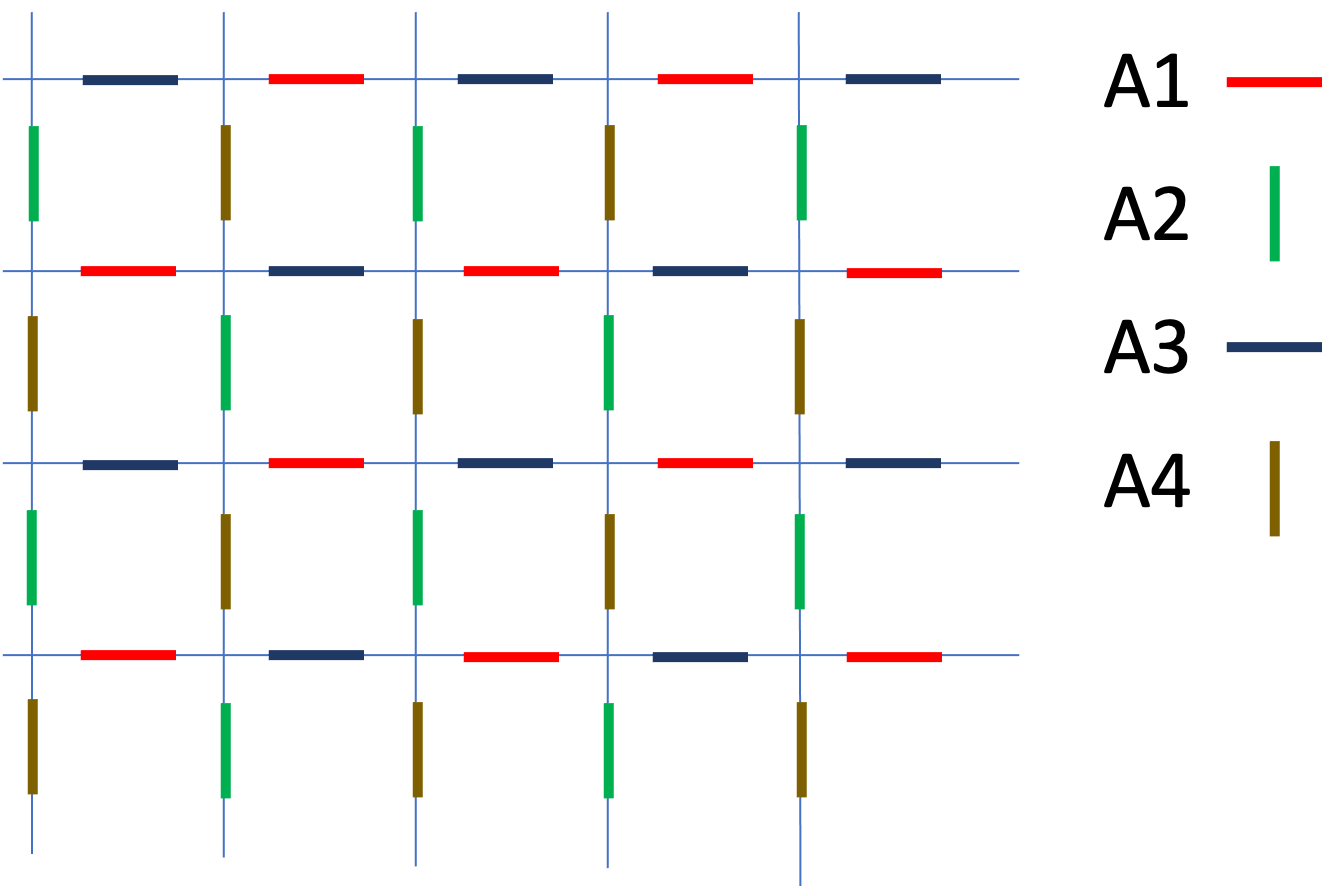}
    \caption{The RLBL model. Hopping is sequentially activated among neighbouring sites connected in the set $A_i$, $i=1,...,4$.}
    \label{fig:RLBL}
\end{figure}

As a particularly illuminating example, consider the Rudner-Lindner-Berg-Levin model \cite{rudner2013anomalous}. This model is an exact toy model for a topological Floquet insulator and has been very useful in flushing out some of their salient properties. In addition, it provides the starting point for other states, such as the anomalous Floquet-Anderson insulators \cite{titum2016anomalous}. The model is two dimensional, however, it's simplicity lies in its similarity to even-odd type models, \cite{Kumar2018evenodd,Ljubotina2019evenodd,Piroli2020evenodd,Lu2022EvenOddPRL}, in that the evolution activates disjoint pairs of sites at each stage. The model can be tuned to a particular point where the stroboscopic evolution of product states is deterministic exhibiting bulk periodic motion and edge propagation. Similarly, one can tune the driving frequency to completely freeze the stroboscopic evolution.
Here, we add interactions to the model and ask when we can make the evolution a product state permutation, at least in some sectors. 
The Hubbard-RLBL evolution is written as 
\begin{eqnarray} \label{eq: Floquet steps}
U=U_{wait} U_4 U_3 U_2 U_1 
\end{eqnarray}
where $U_i(V,\tau)=e^{-i \tau H_i}$. For $i=1,..4$,
\begin{gather}
    {\cal H}_i = -t_{hop}\sum_{(i,j)\in A_i;\sigma}  (a_{i,\sigma}^\dagger  a_{j,\sigma} + h.c.)  + V \sum_{i\in A_i} n_{i,\uparrow} n_{i,\downarrow}
    \label{eq: RLBL Hubbard Hamiltonian}
\end{gather}
where $n_{i,\sigma}=a_{i,\sigma}^\dagger  a_{i,\sigma}$ and the sets $A_i$ are described in Fig. \ref{fig:RLBL}.  

Note, this is equivalent to the model investigated in \cite{Nathan2021AFI} when $U_{wait} \rightarrow U_{dis}$, i.e. the waiting period corresponds to evolution with random local potentials and no hopping \footnote{Technically, in \cite{Nathan2021AFI} a weak disorder potential is added during the $U_i$ steps and then the disorder strength during the wait step is effectively made stronger by increasing the length of time the wait step is applied.  However, this slight difference in how the disorder potential is applied does not seriously alter the dynamics and so we will not make a hard distinction between the two.}.  In that work, it was shown that this model supports a new family of few-body topological phases characterized by a hierarchy of topological invariants.  These results may be viewed from the following perspective. First, finely-tuned points where the dynamics is exactly solvable were studied (namely, $\tau=\frac{\pi}{2}$ and $V=0$ or $V \rightarrow \infty$).  Second, it is argued that regions near these special points are stabilized (i.e. localized, at least for finite particle number cases) by disorder leading to robust phases.  Finally, topological invariants characterizing these phases (V small vs. V large) can be found and shown to be distinct implying two differing topological phases.  An application of the methods we propose in this work will allow us to generalize the first step above and find families of these exactly solvable points.  We leave discussions of when regions in parameter space near these points may or may not be stabilized by disorder to future work.  Since, at these exactly solvable points, we will be mapping product states to product states, $U_{dis}$ will only act as an unobservable global phase and thus for the rest of our analysis we will set $U_{wait} = I$.  Furthermore, throughout the rest of the paper we will work in units where $t_{hop}=1$ and $\hbar=1$.           

%Here $U_{wait}$ allows the system to accumulate additional local phases between the hopping stages. %It is described in \eqref{eq: U wait} and its consequences are detailed in section (\ref{sec:classical ev disorder}). 
%In the evolution we consider here, $U_{wait}$ acts by adding a trivial phase and we can thus take it to be $U_{wait}=I$. 

We now look for conditions to simplify the evolution \eqref{eq: Floquet steps} in such a way that the total evolution reduces to a  permutation on the set of product states, i.e. when an initial configuration of fermions is placed at a selection of locations it will evolve into a different assignment of locations without generating entanglement.

To do so, we note that the evolution of each pair of sites, may be considered separately due to the disjoint nature of the set of pairs $A_i$. Thus, we consider the  evolution on a pair of sites $i,j$
\begin{eqnarray}
U_{(i,j)}(V,\tau)=e^{-i \tau (a_{i,\sigma}^\dagger  a_{j,\sigma} + h.c.)  +\tau V(  n_{i,\uparrow} n_{i,\downarrow}+ n_{j,\uparrow} n_{j,\downarrow}
    \label{eq: RLBL Hubbard pair evolution} }.
\end{eqnarray}
Since the evolution preserves particle number, we can treat the sub-spaces of $0,1,2,3,$ and $4$ particles in each neighboring pair of sites separately.
In the case of $0$ or $4$ particles, evolution is trivially the identity (due to Pauli blocking in the $4$ particle case).  For $1$ or $3$ particles, one of the two sites is always doubly occupied, and thus the interaction term in \eqref{eq: RLBL Hubbard Hamiltonian} is a constant and does not affect evolution. In this case, solving the two site non-interacting evolution we see that in the one-particle sector, a fermion starting initially at site $i$ has a probability $p = \sin^2{ \tau}$ to hop to the other site in pair $j$ and probability $1-p$ to stay.  Similarly, in the 3-particle sector, an initially placed hole in site $i$ has the same probability, $p$ to hop to the other site $j$. Thus, when
\begin{gather}
    \tau = \frac{\pi}{2} \ell 
    \label{eq: 2 site perm condition 0}
\end{gather}
for some integer $\ell$, evolution for initial product states in the $1$,$3$ particle subspace is completely deterministic with trivial evolution for even $\ell$ and the particle hopping to the other site in the pair with probability $1$ (henceforth referred to as perfect swapping) when $\ell$ is odd.  Clearly, for these values of $\tau$ (and independently of $V$), no new entanglement is created in any pairs with $1$ or $3$ particles.  To render the evolution in the $2$ particle pair subspace simple, it is shown in appendix \ref{appendix: hubbard 2 sector} that deterministic evolution occurs when the two conditions below are simultaneously satisfied:
\begin{gather}
    \tau \sqrt{4^2 + V^2} = 2 \pi m \label{eq: 2-site permutation condit 1} \\
    \text{and} \nonumber \\
    \frac{1}{2} \tau V + \pi m = \pi n
    \label{eq: 2-site permutation condit 2} 
\end{gather}
 with $n,m \in \integers$.  Note that \eqref{eq: 2-site permutation condit 1} guarantees the preservation of the number of doubly occupied sites (doublons). When $n$ is even, the sub-system will return to its initial state.  On the other hand, if $n$ is odd, the system will exhibit perfect swapping i.e. each particle will hop to the other site in the pair.  By solving for $\tau$ and $V$ in terms of $n$ and $m$, we may now summarize when evolution is deterministic in each of the particle number sub-spaces:
 
 \begin{gather} \label{eq: freeze condition}
 \begin{tabular}{c | c | c}
    particles & $\tau$ & $V$ \\ \hline
      1 or 3 & $\tau = \frac{\pi}{2} \ell$ & $V$ arbitrary\\
      2, opposite spins & $\tau = \frac{\pi}{2} \sqrt{2 m n - n^2}$ & $V = \frac{4 (n-m)}{\sqrt{2 m n - n^2}}$ \\
      otherwise & any & any
 \end{tabular}
\end{gather}
when $n$ or $\ell$ are even (odd) evolution is frozen (perfect swapping). To keep the solutions real, Eq. \ref{eq: freeze condition} also implies we must take $2 m n-n^2>0$.

Can all the conditions \eqref{eq: 2 site perm condition 0}, \eqref{eq: 2-site permutation condit 1}, and \eqref{eq: 2-site permutation condit 2} be simultaneously satisfied? In such a case the evolution of $\cal U$ is simply a permutation (being a product of identities and site swaps) and generates no new entanglement in any of the sectors. %This requires that $\frac{\pi}{2} \ell = \frac{\pi}{2} \sqrt{2 m n - n^2}$ or

\subsection{The Diophantine Equation}
Combining the conditions \eqref{eq: 2 site perm condition 0}, \eqref{eq: 2-site permutation condit 1}, and \eqref{eq: 2-site permutation condit 2} together yields the following equation:
\begin{gather} \label{eq: diophantine Floquet Hubbard}
    \ell^2 + n^2 = 2 m n. \,\, \ell,n,m \in \integers
\end{gather}
Eq. \eqref{eq: diophantine Floquet Hubbard} is a homogeneous Diophantine equation of degree $2$ and can be solved. 

We now give a brief review of Diophantine equations and the strategy for solving homogenous quadratic equations. 

Diophantine equations are algebraic (often polynomial) equations of several unknowns where only integer or rational solutions are of interest.  They are named in honor of Diophantus of Alexandria for his famous treatise on the subject written in the 3rd century though the origins of Diophantine equations can be found across ancient Babylonian, Egyptian, Chinese, and Greek texts \cite{Cohen2007Dioph}.  Despite their often innocuous appearance, they are an active area of research with solutions frequently requiring surprisingly sophisticated mathematical techniques and have been the centerpiece of several famous, long-standing mathematical problems that have only been (relatively) recently resolved, including Fermat's Last Theorem \cite{Wiles1995Fermat} and Hilbert's Tenth Problem \cite{Matiyasevich1970Hilbert}.

In this section, we are interested in the relatively simple case of a homogeneous quadratic Diophantine equation, i.e. equation of the form

\begin{gather}
    X^T Q X = 0
    \label{eq:quad Diophantine}
\end{gather}
with variables $X^T = \left(x_0,x_1,...,x_n \right)$ and coefficients given by the $n\times n$ symmetric matrix $Q$ with integral diagonal entries and half integral off-diagonal entries.  As we shall see, however, for interactions beyond Hubbard a broader class of Diophantine equations may need to be considered.  For information on broader classes of Diophantine equations and for more information on the derivation to follow, see, for example, \cite{Cohen2007Dioph}.   

The general strategy for finding rational (we will specialize to integer solutions for our cases of interest at the end) solutions to \eqref{eq:quad Diophantine} is to first find a particular solution and then generate all other rational solutions from the particular solution.  Particular solutions can be found simply by inspection or through existing efficient algorithms \cite{Cohen2007Dioph}.  The main task is then to generate all other rational solutions from a given particular solution.

\begin{figure}
    \centering
    \includegraphics[width=0.3\textwidth]{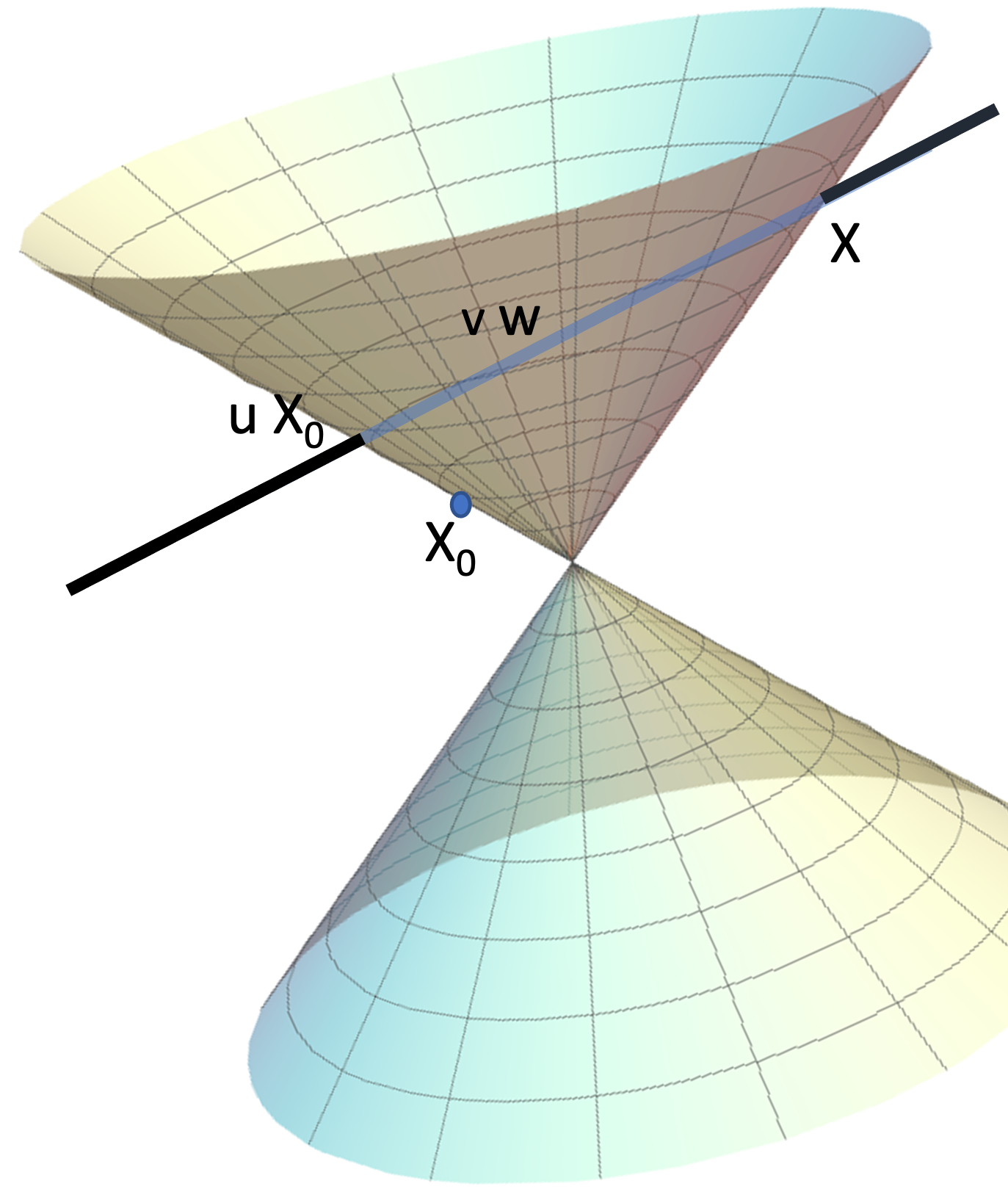}
    \caption{Any line passing through the null surface has two points of intersection. Given a particular solution $X_0$ of the homogeneous Diophantine eq  \eqref{eq:particular quad Diophantine}, other rational solutions are found by looking at lines emanating from $u X_0$ with rational slopes .}
    \label{fig:Diophantine}
\end{figure}

Take $X_0^T = \left(x_{0,0},x_{1,0},...,x_{n,0} \right)$ to be a particular solution, i.e. 

\begin{gather}
    X_0^T Q X_0 = 0.
    \label{eq:particular quad Diophantine}
\end{gather}
Since \eqref{eq:quad Diophantine} is quadratic, any line through $X_0$ will intersect the hypersurface defined by \eqref{eq:quad Diophantine} at a single other point (see Fig. \ref{fig:Diophantine}).  Furthermore, if the line through $X_0$ is rational (i.e. has rational coefficients), as we see below, this implies that the second intersection point must also be rational.  Therefore, it is possible to generate every rational solution to \eqref{eq:quad Diophantine} by finding the second intersection point of every rational line through $u X_0$, where $u$ is rational.

Here, since \eqref{eq:quad Diophantine} is homogeneous, it is convenient to work in projective space $\mathbb{P}_n(\mathbb{Q})$ where a general line passing through $X_0$ is parameterized by 

\begin{gather}
    X = u X_0 + v W
    \label{eq:gen line through X0}
\end{gather}
with $(u,v) \in \mathbb{P}_2(\mathbb{Q})$ and any $W=(w_1,..,w_n) \in \mathbb{P}_n(\mathbb{Q})$ not equal to $X_0$.  Combining \eqref{eq:gen line through X0} and \eqref{eq:quad Diophantine},

\begin{gather}
    0 = (u X_0 + v W)^T Q (u X_0 + v W) \\
    = v \left(2u W^T Q X_0 + v W^T Q W \right)
\end{gather}
where we have simplified using \eqref{eq:particular quad Diophantine}.  We may thus take as the solution $(u,v) = \left(W^T Q W,-2 W^T Q X_0 \right)$.  Combining with Eq \eqref{eq:gen line through X0} and multiplying by a general $d \in \mathbb{Q}$ to restore full solutions (since we considered $X$ as an element of a projective space), we find

\begin{gather}
    X = d \left[(W^T Q W)X_0 - 2 (W^T Q X_0) W \right].
    \label{eq:final X}
\end{gather}
For integer solutions, we need simply to rescale $W \rightarrow \frac{W}{\zeta}$ and $d \rightarrow d \zeta^2$ where $\zeta = \gcd({w_i})$.  After rescaling, the only non-integer information is coming from $d$, so all integer solutions may be found simply by considering integer $d$.

For the relevant case of $n=3$, let us, without loss of generality, diagonalize $Q = diag(A,B,C)$ and let $W^T = (w_1,w_2,0)$ where (after rescaling with $\zeta$) $w_1$ and $w_2$ are co-prime integers and the final element of $W$ may be set to $0$ due to the required linear independence with $X_0$.  Simplifying \eqref{eq:final X} then becomes

\begin{gather}
    X = d(A w_1^2 + B w_2^2) \left(\begin{tabular}{c}
          $x_{0,0}$ \\
          $x_{1,0}$  \\
          $x_{2,0}$ 
    \end{tabular}\right)
    \nonumber \\
    - 2d (w_1 A x_{0,0} + w_2 B x_{1,0}) \left(\begin{tabular}{c}
          $w_1$ \\
          $w_2$  \\
          $0$ 
    \end{tabular}\right) \\
    = d \left(\begin{tabular}{c}
         $-(A w_1^2 - B w_2^2 )x_{0,0} - 2 B w_1 w_2 x_{1,0}$  \\
         $(A w_1^2 - B w_2^2 )x_{1,0} - 2 A w_1 w_2 x_{0,0}$\\
         $(A w_1^2 + B w_2^2 )x_{2,0}$
    \end{tabular} \right)
    \label{eq: final 3 quadratic dio solution}
\end{gather}

\subsection{Solution for product state permutation dynamics with Hubbard interaction}
Following the previous section, we write our Diophantine eq. \eqref{eq: diophantine Floquet Hubbard} in a diagonal form:
\begin{gather}
    \ell^2 + n^2 = 2m n\\
%    \implies \ell^2 + \Tilde{n}^2 - m^2 = 0 \\
    \implies \left(\begin{tabular}{c c c}
         $\ell$ & $\Tilde{n}$ & $m$ \\
    \end{tabular}\right) \left(\begin{tabular}{c c c}
         1 & 0 & 0  \\
         0 & 1 & 0  \\
         0 & 0 & -1  \\
    \end{tabular}\right)
    \left(\begin{tabular}{c}
         $\ell $  \\
         $\Tilde{n}  $ \\
         $m$   \\
    \end{tabular}\right) 
    = 0,
\end{gather}
where we have defined $\Tilde{n} \equiv n-m$.  Note, this is the famous Diophantine equation for Pythagorean triples.

By inspection, a non-trivial solution is $\ell=-1,\Tilde{n}=0,m=1$.  Utilizing Eq. \eqref{eq: final 3 quadratic dio solution} we find

\begin{gather}
    \left(\begin{tabular}{c}
         $\ell$  \\
         $\Tilde{n}$ \\
         $m$
    \end{tabular}\right) = d \left(\begin{tabular}{c}
         $w_1^2 - w_2^2$  \\
         $ 2 w_1 w_2$\\
         $ w_1^2 + w_2^2$
    \end{tabular} \right) \label{eq:pythag triples} \\
    \implies \left(\begin{tabular}{c}
         $\ell$  \\
         $n$ \\
         $m$
    \end{tabular}\right) = d \left(\begin{tabular}{c}
         $w_1^2 - w_2^2$  \\
         $[w_1+w_2]^2$\\
         $ w_1^2 + w_2^2 $
    \end{tabular} \right)
    \label{eq:final hubbard dio solution}
\end{gather}
Note, Eq. \eqref{eq:pythag triples} is the standard solution for Pythagorean triples.

We thus found that the set of  $n$, $m$, and $\ell$ simultaneously satisfying the conditions for simple dynamics can be written as: 
\begin{subequations} \label{eq: diophantine solution}
\begin{gather}
    \ell = d(w_1^2 - w_2^2) \\
    m = d(w_1^2 + w_2^2) \\
    n = d(w_1 + w_2)^2
\end{gather}
\end{subequations}
where $d,w_1, w_2 \in \integers$ and $w_1,w_2$ are coprime.  Note, in \eqref{eq: diophantine solution}, if $\ell$ is even (odd) then so is $n$.  This implies that the only way to completely satisfy the conditions in Eq. \eqref{eq: freeze condition} is if all motion is frozen or all motion (not constrained by Pauli exclusion) becomes perfect swapping.  

Inspecting the above solutions, we see that $2 m n-n^2=(w_1^2-w_2^2)^2$, automatically satisfying the condition $2 m n-n^2>0$ for $V$ and $\tau$ to be real. 
Finally our solution is summarized by
\begin{eqnarray}
\tau=\frac{\pi}{2} d(w_1^2 - w_2^2)\,\,;\,\, V=\frac{8 w_1 w_2}{|w_1^2-w_2^2|}.
\label{eq: tau and V hubbard}
\end{eqnarray}
Note that $V$ doesn't depend on the choice of $d$, and that any choice involving $w_1=0$ or $w_2=0$ will yield a non-interacting model.
As an illustration, consider the following example choices:
\\
1. Taking $w_1=1,w_2=0,d=1$ yields $\tau={\pi \over 2},V=0$, which is the non-interacting dynamics considered in the original RLBL model, with perfect swapping. 
\\
2. Taking $w_1=3,w_2=1,d=1$ yields $\tau={4 \pi },V=3$.  Since $\ell$ is even in this case, the dynamics is completely frozen.
\\
3. Taking $w_1=3,w_2=-1,d=1$ yields $\tau={4 \pi },V=-3$, i.e. frozen dynamics in a model with an attractive Hubbard interaction.

It is important to note that the special values of interaction strength and driving frequency in Eq. \eqref{eq: tau and V hubbard} hold for any Hubbard-Floquet procedure where hopping between pairs of sites is sequentially activated.  This is the case for such systems on any lattice and in any dimension. 
We also note, that the Diophantine solution is ill suited to describe the singular case of infinite $V$ and finite $\tau$ and therefore this situation must be handled separately.  In the limit of large $V$, the interaction strength overpowers the hopping strength and all evolution is frozen in the 2-particle sector.  On the other hand, evolution in the 1,3 particle sector is independent of $V$ and therefore may exhibit perfect swapping or freezing.  Thus, in this case, it is possible to have one sector (the 2-particle sector) frozen while the other (the 1,3 particle sector) exhibits perfect swapping.

\subsection{Example 2: Nearest neighbour interactions on a Lieb lattice.}
\label{section: NN RLBL model}
\begin{figure}
    \centering
    \includegraphics[width=0.4\textwidth]{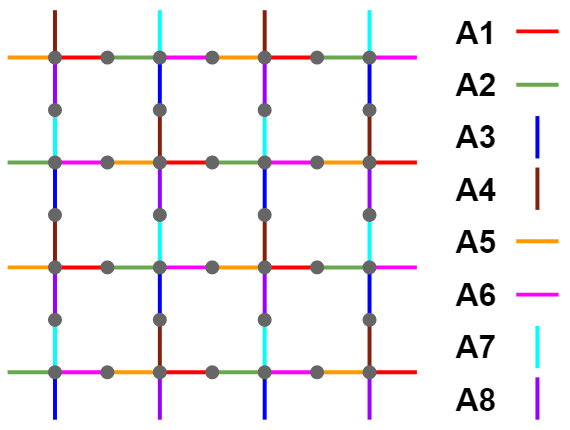}
    \caption{RLBL-like model on a Lieb lattice.  Hopping between neighboring pairs of sites within $A_i$ is activated during step $i$ of the Floquet drive.  The same sequence of activated site pairs is achieved with the chiral measurement scheme introduced in \cite{wampler2021stirring}.   During each step $i$, evolution is confined between neighboring sites in $A_i$ by rapidly measuring (in the Zeno limit) all sites in the complimentary set $A_i^c$.  Both models, with NN interactions, will share the same conditions (Eqs. \eqref{eq: NN delta not 0 condit} and \eqref{eq: NN delta 0 condit}) for number state to number state evolution.}
    \label{fig:MIC}
\end{figure}

In the next two examples, we consider interactions involving nearest neighbours.  Unfortunately, adding nearest neighbour interactions to the RLBL model directly destroys an essential feature for the solvability of the problem: that the evolution operators of different pairs of sites are not directly coupled (and therefore commute). Here, instead, we choose to work with RLBL-like dynamics on a Lieb lattice as described in \cite{wampler2021stirring}.  The dynamics we consider here essentially activates pairs that are separated by several lattice sites at each step. The sequence of activations is described in Fig \ref{fig:MIC}.

\begin{figure}
    \centering
    \includegraphics[width=0.2\textwidth]{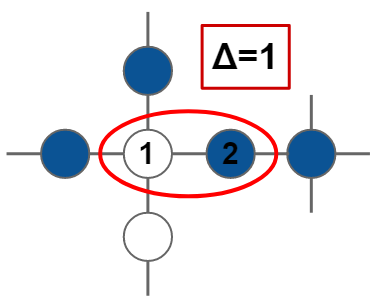}
    \caption{Evolution of a 2-site pair in the NN-RLBL model on a Lieb lattice.  All evolution is restricted to the red ellipse above.  Evolution within the red ellipse (i.e. between site 1 and site 2) is determined by $\tau$, $V$, and the neighboring particle number difference $\Delta = |N_1 - N_2|$.  In this case, $N_1 = 2$ and $N_2 = 1$, so $\Delta = 1$.  If the $\Delta = 1$ condition on $V$ and $\tau$ in Eq. \eqref{eq: NN delta not 0 condit} is satisfied, then the particle at site 2 will exactly return to site 2 after a time $\tau$ (at intermediate times, the particle may be in a generic superposition of being located at site 1 and site 2).  }
    \label{fig:Delta}
\end{figure}

Here, we consider spinless fermions on the Lieb lattice. There are 8 steps. At step $i$ we activate hopping between sites that are nearest neighbours that belong to the set $A_i$. The evolution is given by:
\begin{gather}
    U= U_8 U_7 U_6 U_5 U_4 U_3 U_2 U_1 
    \label{eq: NN-RLBL U}
\end{gather}
where $U_i = e^{-i {\cal H}_i \tau}$, and
\begin{gather}
    {\cal H}_i = -t_{hop}\sum_{(i,j)\in A_i }  (a_{i }^\dagger  a_{j } + h.c.)  +  V \sum_{<i,j>} n_i n_j
    \label{eq: Lieb Hamiltonian}
\end{gather}
We proceed, as in Section \ref{Section: Hubbard RLBL}, by considering the evolution of a single connected pair during step $i$ and exactly solving for values of $V$ and $\tau$ where the pair exhibits freezing or perfect swapping.  The evolution of a 2-site pair of sites $i,j$ for one step is given by
\begin{gather}
U_{(i,j)}=e^{-i\tau[
 - t_{hop}( a_{i}^\dagger a_{j} + h.c.) + V n_i\sum_{k:\langle i,k\rangle} n_{k}+V n_j\sum_{k:\langle j,k\rangle}  n_{k}]}.
 \label{eq: 2 site NN evolve}
\end{gather}
%Below, we will find conditions for Fock state evolution under \eqref{eq: 2 site NN evolve}. 
Note that the number operators on neighbours of $i,j$ commute with the evolution.
Let the initial number of occupied neighbours of the sites $i$ and $j$ be $N_i$ and $N_j$ respectively (not counting $i,j$ themselves).  Evolution of the 2-site pair is now exactly solvable in terms of $\Delta=N_i-N_j$, the difference in the number of particles neighboring sites $i$ and $j$ in the 2-site pair respectively (see Figure \ref{fig:Delta}). 

Solving the two site evolution, we find that evolution is frozen when 
 \begin{eqnarray}\label{eq:DeltaEq}
 \sqrt{4 + \Delta^2 V^2} \tau = 2 \pi m
  \end{eqnarray}
  for some $m \in \integers$. We find that the evolution may only be perfect swapping when $\Delta=N_i-N_j = 0$ and occurs when $\tau = \frac{\pi}{2} + \pi m$ for $m \in \integers$ (see appendix \ref{appendix NN RLBL} for details).

  In the rest of the paper, whenever considering the evolution on a pair of sites, we will denote $\Delta$ as the difference in the number of (static) particles that are nearest neighbours of the two sites during the relevant evolution step.

\subsection{A coupled set of Diophantine Equations}
\label{Section: coupled diophantine}
For a generic initial position of the particles, $N_i-N_j$ will not be uniform across the sample. Thus, for proper particle permutation dynamics, we must simultaneously find a solution of \eqref{eq:DeltaEq} for all possible values of $|N_i-N_j|$. 

Note that $N_i$ takes the values $0,..,D_i-1$, where $D_i$ is the degree (number of neighbours) of lattice site $i$. It follows that   $|N_i-N_j| \in\{0,..,max(D_i,D_j)-1\}$. Thus, if $D_{max}$ is the maximum degree of the lattice, we have the simultaneous conditions:
\begin{gather}
    \sqrt{4 + \Delta^2 V^2} \tau = 2 \pi m_{\Delta} \hbox{ } \forall \hbox{ } \Delta =1,...,(D_{max}-1) \label{eq: NN delta not 0 condit} \\
     \tau = \frac{\pi}{2} m_0 \hbox{ } \text{corresponds to} \hbox{ } N_i=N_j
    \label{eq: NN delta 0 condit}
\end{gather}
with all $m_i \in \integers$.

Equations \eqref{eq: NN delta not 0 condit} and \eqref{eq: NN delta 0 condit} provide $D_{max}$ equations that must be solved simultaneously.  The first two equations set the values for $\tau$ and $V$ in terms of $m_0$, $m_1$:  
\begin{eqnarray}\label{eq:tauAndV}
\tau=\frac{\pi}{2} m_0\,\,;\,\, V^2=4( \frac{4m_1^2 }{m_0^2}-1).
\end{eqnarray}
However, the rest of the equations for $m_i$, with $i>1$, must be simultaneously solved with these values for $\tau$ and $V$ yielding the coupled equations: 
\begin{gather}\label{eq:NN Diophantine set}
    4m_l^2=(1-l^2 ) m_0^2+4 l^2 m_1^2\,\\ \,m_l\in \integers \, , \, l =2,3,...,(D_{max}-1)
\end{gather}
A first solution to this system may be obtained by taking $m_0=2m_1=2m_2=...=2m_{D_{max}-1}$, which, by \eqref{eq:tauAndV}, yields the non-interacting case $V=0$.  We now search for non-trivial solutions (i.e. $\tau,V \neq 0$). 

{\it Solution for $D_{max}=3$}. For $D_{max}=3$, we describe a general solution in appendix \ref{appendix NN RLBL} that yields non-trivial solutions.  The result:  
\begin{gather}
    \left(\begin{tabular}{c}
         $m_0$ \\
         $m_1$\\
         $m_2$
    \end{tabular} \right) = d \left(\begin{tabular}{c}
         $ - 32 w_1 w_2$  \\
         $-3 w_1^2 - 16 w_2^2  $\\
         $2 \left[-3 w_1^2 + 16 w_2^2 \right]$
    \end{tabular} \right)
    \label{eq:NN Dio Solution 0 1 2}.
\end{gather}
We note that $m_0$ is always even and thus all evolution is frozen.  Due to the hierarchy of the equations, total freezing must then occur for any solutions with $D_{max} \geq 3$.  

{\it Solution for $D_{max}=4$}. We  combine equations \eqref{eq:NN Dio Solution 0 1 2} and the $\Delta=3$ equation from \eqref{eq: NN delta not 0 condit} to find a new Diophantine equation for the case $D_{max}=4$:
\begin{gather}
    m_3^2 = 81 w_1^4 + 2304 w_2^4 - 1184 w_1^2 w_2^2
    \label{eq: NN 4 delta Dio}
\end{gather}
The Diophantine equation \eqref{eq: NN 4 delta Dio} is harder to solve. However, a numerical search does find non-trivial ($V \neq 0$) solutions.  For example, $(w_1;w_2;m_3) = (3;9471;4305592257)$ is a solution with $V \approx 6,394$ and $\tau = 454,608 \pi$.  Whether there exist $V,\tau$ such that lattices with a maximum degree larger than $4$ may exhibit fully product state permutation evolution is an open question. 
The result for $D_{max}=4$ suggests the conjecture that there are solutions to the system of equations for any $D_{max}$. Similar to the strategy above, by solving for $D_{max}=k$, it is possible to construct a new Diophantine equation for $D_{max}=k+1$. Determining whether  this tower of equations is solvable is outside the scope of the present paper.  On the other hand, as can already be seen in the case of $D_{max}=4$, the values of $V,\tau$ for which the system exhibit such freezing for any initial number state quickly become prohibitively large for typical physical systems as the maximum lattice degree increases.  
%On the other hand, states that evolve with CA may still exist in lattices with higher max degrees and reasonable $V,\tau$ when only 2 or 3 of the equations \eqref{eq:NN Vt restriction} are satisfied as discussed in Section \ref{Section: Scars}.    For $D_{max} = 4$, as is the case for the Lieb lattice being considered here, we numerically find that non-trivial solutions exist for the coupled Diophantine equations \eqref{eq:NN Diophantine set}.  It is an open question whether non-trivial solutions exist for $D_{max}>4$, however as $D_{max}$ increases the values of $V,\tau$ that satisfy the increasing number of coupled Diophantine equations quickly become large and unreasonable on experimental grounds.  
\old

{\it Remark.} It is straightforward to generalize the Hamiltonian \eqref{eq: Lieb Hamiltonian} to include more elaborate interactions as long as at each step the number operators associated with the neighbourhood of each evolving pair is constant. For example, we can write
\begin{gather}
    {\cal H}_i =  -t_{hop}\sum_{(i,j)\in A_i}  (a_{i}^\dagger  a_{j} + h.c.)  +  \sum_{i\in A_i} V_{ij}n_{i} n_{j} ,
    \label{eq: RLBL general}
\end{gather}
Given the number of particles in the neighborhood of each 2-site pair, we write (note here we include the potentials $V$ in the the definition of $\Delta$):
\begin{eqnarray}
\Delta_{ij}= \sum_{k:\langle i,k\rangle} V_{ik}n_{k}- \sum_{k:\langle j,k\rangle} V_{jk} n_{k}\label{eq: Delta forms}
\end{eqnarray}
and the freezing condition becomes:
\begin{gather}
    \tau \sqrt{4 + \Delta_{ij}^2 } = 2 \pi m_{ij},  \,\, m_{ij} \in \integers
\end{gather}
for all $\Delta_{ij}$ of the form \eqref{eq: Delta forms}.

%\section{Scar-like states in a Hybrid system}\label{Section: Scars}

\subsection{Example 3: Deterministic evolution in the measurement induced chirality model on a Lieb lattice.}
\label{section: MIC}

As another example, We consider the measurement induced chirality protocol of   \cite{wampler2021stirring} with added nearest neighbour interactions and in the Zeno limit. In that work, a simple hopping Lieb lattice model of fermions was subjected to repeated measurements changing according to a prescribed chiral protocol. In contrast to the previous models, the Hamiltonian is not time dependent and all hopping terms in the Hamiltonian remain activated throughout the process. 

It was shown in  \cite{wampler2021stirring}  that in the limit of rapid measurements, the so called the Zeno limit, the resulting dynamics is a classical stochastic process of permuting Fock states. We will see that, in this case too, we can find special values of interaction strength and protocol duration where the dynamics becomes deterministic. In fact, we will see the dynamics is governed by the same Diophantine equation as in example 2. 

%We will see that it is also possible to find points of deterministic permutative evolution in systems with static, interacting Hamiltonians where periodic procedures of measurements are applied that utilize the quantum Zeno effect to sequentially isolate disjoint pairs of sites.  Evolution of this type is of recent interest. For example, the work \cite{wampler2021stirring} has shown that RLBL-like evolution can be induced in static systems solely through a measurement protocol of this type.  Similar to RLBL, the measurement induced dynamics supports protected chiral charge transport along the edge of the system simultaneously with zero bulk transport.  The system also exhibits new phenomena due to the non-unitary nature of the evolution.  

  %We find the corresponding Diophantine equations and CA evolution of the system in order to exemplify how the techniques from Section \ref{Section: Hubbard RLBL} may be used in systems with other interactions, on other lattices, and with measurement-induced instead of Floquet procedures.  
  
%In this section, we analyze the measurement protocol on a Lieb lattice originally introduced in \cite{wampler2021stirring} with added nearest-neighbor interactions.
Specifically, we consider fermions hopping on a Lieb lattice with nearest-neighbor interactions given by
\begin{gather} \label{eq: H NN}
    {\cal H} = -t_{\text{hop}} \sum_{<i,j>} a_i^\dagger a_j + V \sum_{<i,j>} n_i n_j .
\end{gather}
We now apply the measurement protocol introduced in \cite{wampler2021stirring} to the system.  Namely, we consider an 8 step measurement protocol in which, during the $i^{th}$ step that runs for a time $\tau$, the local particle density in all sites in a set $A_i^c$ of sites are measured. In the Zeno limit, all evolution during a step is restricted to neighboring sites in the subspace $A_i$ (See figure \ref{fig:MIC} for details), while the rest of the sites are kept frozen. Thus, in the Zeno limit, the evolution is effectively split into $8$ steps evolved by the Hamiltonian \eqref{eq: Lieb Hamiltonian}, interspersed by an additional measurement. The measurements keep projecting the system onto Fock states, however, the particular states at hand are statistically distributed. However, if the step evolution \eqref{eq: 2 site NN evolve} maps Fock states into Fock states, the whole procedure yields a deterministic evolution of an initial Fock state into another.  In other words, the conditions for permutative evolution (and the corresponding set of Diophantine equations) for this model are equivalent to those found in the interacting Floquet model investigated in example 2.

\section{Hilbert Space Fragmentation}\label{Section: Scars}
In Section \ref{Section: coupled diophantine}, we gave $D_{\text{max}}$ conditions that must be simultaneously satisfied for Fock state permutative dynamics in models on a Lieb lattice with NN interactions.  Similarly, in Section \ref{Section: Hubbard RLBL} we gave conditions for permutative evolution in the Floquet-Hubbard RLBL model.  If in these models not all of these conditions are satisfied, then the evolution of a general initial state will require consideration of the full quantum many-body Floquet Hamiltonian.

However, evolution for certain initial states may still be deterministic even if only one or a few of the conditions for Fock state to Fock state evolution are met.  This fragments \cite{Moudgalya2022Scars} the Hilbert space, ${\cal H}$, into disconnected Krylov supspaces, ${\cal K}_i$, i.e.

\begin{gather}
    {\cal H} = \bigoplus_i {\cal K}_i ,~ {\cal K}_i = span_n\{{\cal U}^n |\psi_i \rangle \}
\end{gather}
where we have chosen a states $|\psi_i\rangle$ that are number local states in such a way that ${\cal K}_i$ are unique.  In the rest of this section, we will explore the nature of the Hilbert space fragmentation in the example interacting Floquet and measurement induced models discussed in the previous section.  Namely, we will see how the Hilbert spaces in these systems simultaneously support Krylov subspaces that are one-dimensional and correspond to frozen product states, few dimensional and correspond to states that evolve according to a classical cellular automation \cite{Wolfram1983CA}, and exponentially large subspaces that may evolve with more generic quantum many-body evolution.    

\subsection{Arrested development}

\begin{figure}[h]
    \centering
    \includegraphics[width=0.3\textwidth]{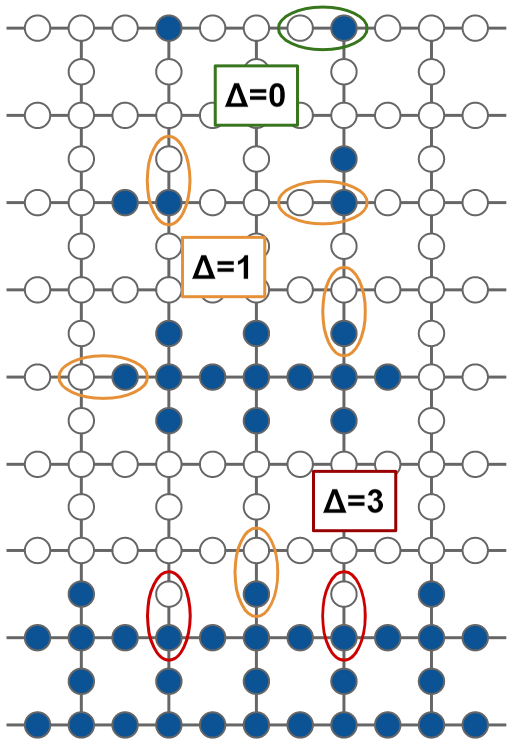}
    \caption{A Zoo of frozen particle configurations when only some of the conditions in \eqref{eq: NN delta not 0 condit} and \eqref{eq: NN delta 0 condit} are satisfied on a nearest neighbour interacting Lieb-RLBL model.  At the top, a particle configuration that requires only that the $\Delta=0$ condition (and $m_0$ even) be satisfied for frozen evolution. %(if $m_0$ odd, the particles would move chirally along the edge as in the non-interacting RLBL model).
    In the bulk of the system are particle configurations that will be frozen so long as the $\Delta = 1$ condition is satisfied.  The lower edge of the system provides an example of a particle configuration that will be frozen so long as both the $\Delta=1$ and $\Delta=3$ conditions are satisfied.  Since all the particle configurations above are disjoint, the simultaneous satisfaction of the $\Delta = 0$, $\Delta = 1$, and $\Delta=3$ conditions implies that the entire system above will be frozen.  Each frozen particle configuration corresponds to a 1D Krylov subspace of the full Hilbert space.}
    \label{fig:CA zoo}
\end{figure}

Let us take as an example the NN-RLBL model on a Lieb lattice considered in Section \ref{section: NN RLBL model}.  We have seen that satisfying the i$^{th}$ condition in equations \eqref{eq: NN delta not 0 condit} and \eqref{eq: NN delta 0 condit} implies that evolution on any neighboring pair of sites will be frozen if $\Delta = i$ (and also requiring $m_0$ is even if $\Delta = 0$).  Thus, any given number state will be frozen under the evolution so long as every neighboring 2-site pair in the system containing a single particle has $\Delta=i$.  In figure \ref{fig:CA zoo}, we provide example frozen states for several values of $\Delta$.

Since these frozen states are trivially mapped back onto themselves (stroboscopically), they correspond to one-dimensional Krylov subspaces.  Note, disjoint unions of frozen particle configurations will also be frozen.  Therefore, since the number of possible disjoint unions of these frozen particle configurations grows exponentially with the system size, so too will the number of one-dimensional Krylov subspaces.  

Furthermore, if several of the conditions \eqref{eq: NN delta not 0 condit} and \eqref{eq: NN delta 0 condit} are satisfied, say the $\Delta=i$ and $\Delta=j$ conditions, then a zoo of frozen particle configurations emergres.  Any disjoint unions of particle configurations satisfied by $\Delta=i$ or $\Delta=j$ alone will be satisfied.  Additionally, new frozen particle configurations will emerge that simultaneously require both the $\Delta=i$ and $\Delta=j$ conditions to be frozen.  An example particle configuration requiring the simultaneous satisfaction of the $\Delta = 1$ and $\Delta = 3$ condition is also given in Figure \ref{fig:CA zoo}.   

We emphasize here that the chiral nature of the Floquet procedure played no role in the emmergence of these frozen states.  In fact, any procedure that sequentially activates hopping between neighboring pairs of sites (suitably spaced to keep evolution disjoint after adding NN interactions) will exhibit the exact same frozen states. 

For example, consider a new procedure where, at each step in the evolution, the system is evolved with a $U_i$ from equation \eqref{eq: NN-RLBL U} chosen at random (uniformly), i.e. an example realization of this aperiodic, random evolution is given by

\begin{gather}
\label{eq: chaotic NN-RLBL}
    U = ...U_4 U_5 U_3 U_3 U_1 U_2 U_7 U_3 .
\end{gather}

The exact same states will be frozen in this model as in the NN-RLBL model on a Lieb lattice and therefore the two models will share each of the one-dimensional Krylov subspaces.  In the random model, the Hilbert space fragmentation will thus be split amongst an exponentially large number of one-dimensional, frozen, Krylov subspaces and a single Krylov subspace whose dimension scales exponentially with the system size.  Due to the random, aperiodic nature of the full evolution, we expect evolution within the large-dimensional Krylov subspace to correspond to chaotic dynamics.  In other words, we expect that time evolution of a random initial product state under \eqref{eq: chaotic NN-RLBL} will result in either frozen evolution or ergodic dynamics within the Krylov subspace (for more on Krylov-restricted thermalization see \cite{Moudgalya2021scarbook}).  Both results will occur a finite fraction of the time depending on the initial product state. 

\subsection{Krylov Subspaces of Cellular Automation}\label{section: Cellular Automation}
Since the dynamics of a particle configuration that obey the Diophantine conditions depends crucially on particles on the neighbouring sites,  it can be naturally encoded as a cellular automation step. 
We will now see how Krylov subspaces supporting classical CA \cite{Wolfram1983CA} at each evolution step may emerge in interacting Floquet and measurement-induced systems when a few of the conditions for number state to number state evolution are satisfied.

To elucidate this effect, we consider again the NN-RLBL model on the Lieb lattice.  In this case, we take the $\Delta = 0$ and the $\Delta = 1$ conditions for number state to number state evolution to both be satisfied, but this time the $\Delta=0$ condition is satisfied for perfect swapping while the $\Delta=1$ condition is satisfied for freezing.  This may happen at, for example, $\tau = \frac{\pi}{2}$ and $V=\sqrt{12}$.  

\begin{figure}[h]
    \centering
    \includegraphics[width=0.4\textwidth]{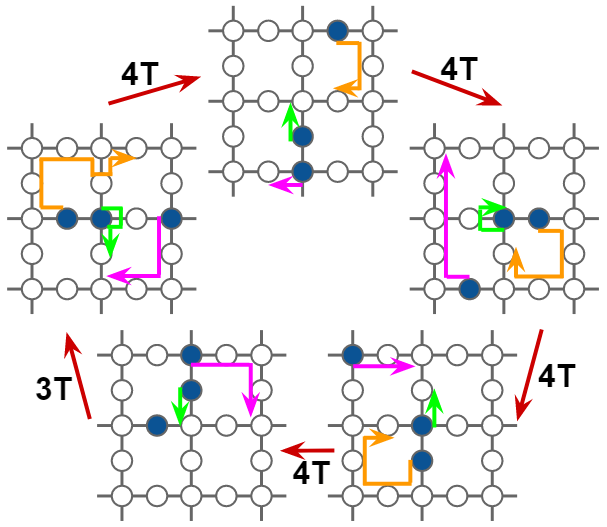}
    \caption{Example evolution within a cellular automation Krylov subspace set by the simultaneous satisfaction of the $\Delta = 0$ and $\Delta = 1$ conditions in equations \eqref{eq: NN delta not 0 condit} and \eqref{eq: NN delta 0 condit}.  In this case, 2-site pairs with $\Delta=0$ evolve with perfect swapping while 2-site pairs with $\Delta=1$ are frozen.  The resulting cellular automation for this example initial particle configuration results in the particles returning to their initial sites after $19T$. Example values of $V,\tau$ that achieve this evolution are $V=\sqrt{12}$ and $\tau = \frac{\pi}{2}$.  Particle trajectories are drawn with orange, green, and magenta arrows.}
    \label{fig: NN 3 bound particles}
\end{figure}

It is now possible to find number states such that the initial particle configuration, $|\Psi_{init}\rangle$, and the resulting states after evolution of each step in the Floquet drive, all satisfy either $\Delta=0$ or $\Delta=1$ for every activated two-site pair in the system with a single particle.  We give an example particle configuration where this may occur in Figure \ref{fig: NN 3 bound particles}.  Here, the space of states $span_n \{U^n |\Psi_{init} \rangle \}$ defines a Krylov subspace where evolution is completely given by a CA since at each step in the Floquet drive the local particle densities are updated deterministically based on the neighboring particle densities (i.e. if $\Delta = 0$ or $1$).        

Similarly to the case of frozen initial particle configurations, disjoint unions of particle configurations that evolve as a CA will also evolve as a CA.  For particle configurations whose CA evolution leaves all particles contained in a volume that does not scale with system size (for example, the evolution of the configuration in Figure \ref{fig: NN 3 bound particles} remains contained within the $5 \times 5$ site square), the number of CA Krylov subspaces will grow exponentially with the system size (since there are exponentially many disjoint unions of such particle configurations).  These CA subspaces may coexist with frozen Krylov subspaces as well as with exponentially large subspaces with more general quantum evolution.  

It is important to note that these CA subspaces break the underlying $T$ time translation symmetry of the evolution operator.  For example, the particle configuration in Fig. \ref{fig: NN 3 bound particles} returns to its initial configuration after $19T$.  However, the exact realization of this Krylov subspace requires fine-tuning in parameter space.  If an alteration of this model was possible such that the realization of these Krylov subspaces did not require fine-tuning, then such a model would be a realization of a time-crystal.  In fact, since the systems we've considered may simultaneously support Krylov supspaces that break the $T$ time translation symmetry in different ways, such a stabilized system would simultaneously support several different time crystals depending on which Krylov subspace contains the initial state.  Recent works \cite{Nathan2019AFI,Nathan2021AFI} have argued that disorder may stabilize dynamics for regions in parameter space near  similarly fine-tuned points in an interacting, Floquet model to acheive anomalous Floquet insulating phases.  We plan to address when disorder may stabilize dynamics for the entire system or for specific Krylov subspaces in our more general set of finely-tuned points in a future work.    

\old

\subsection{Frozen states of Floquet evolution on a chain with nearest neighbour interactions}
A major tool used in the analysis of the interacting Floquet and measurement models above was that the interactions preserved the disjoint nature of the steps of the periodic drive.  However, using the same tools as in the disjoint case, it is possible to find frozen states even when the activated neighboring pairs interact (i.e. do not commute).  

Here, we investigate an example model where the interactions ruin the disjoint nature of the Floquet drive and show how, at special values of interaction strength and driving frequency, it is still possible to find states that are frozen.  Namely, we take as an example a 1D, NN interacting Hamiltonian of the form

\begin{gather}
    {\cal H}(t) = H_0(t) + V\sum_{i=0}^{N-2} n_i n_{i+1}
\end{gather}
where
\begin{gather}
    H_0(t) = 
    \begin{cases}
    \sum_{i \text{ even}}  (a_{i}^\dagger a_{i+1} + h.c.)  & 0 \leq t<\frac{T}{2} \equiv \tau  \\
    \sum_{i \text{ odd}}  (a_{i}^\dagger a_{i+1} + h.c.) & \frac{T}{2} \leq t<T .
    \end{cases}
\end{gather}

%Even-Odd Floquet models such as this are popular toy models whose relative simplicity provides more numerical tractability and, in some cases, exact solvability while simultaneously exemplifying the characteristic, uniquely Floquet properties of periodically driven systems.  For example, variations on this Even-Odd type Floquet model have been shown to exhibit Floquet symmetry protected topological phases \cite{Kumar2018evenodd}, ballistic spin transport \cite{Ljubotina2019evenodd}, and both theoretical and experimental work on a Su-Schrieffer-Heeger type variation has been shown to exhibit Dirac bands \cite{Lu2022EvenOddPRL}.  They also play a central role in investigations into minimal models of quantum chaos via local random unitary circuits where a combination of a lack of conservation laws and the ability to calculate quantities such as Out-of-time-order correlators and entanglement entropy provide fertile ground for study [].  This interest has also inspired more general investigations into when these Even-Odd type models are exactly solvable \cite{Piroli2020evenodd}.  Using similar methods to the previous 3 Sections, We show how, even in some cases where the Even-Odd model isn't exactly solvable, it is still possible to find an extensive number of eigenstates (via the classically evolving states) at special parameter values.

Similarly to the previous cases, let us again consider a single 2-site pair where hopping is activated.  If the occupancy of the sites neighboring the pair happen to be static, then the conditions for frozen or perfect swapping \eqref{eq: NN delta not 0 condit} and \eqref{eq: NN delta 0 condit} will still hold (except here with $D_{max} = 2$).  However, this is, of course, not generally the case.  Even if a neighboring pair is stroboscopically frozen, this is already enough to make $\Delta$ ill-defined and ruin the conditions \eqref{eq: NN delta not 0 condit} and \eqref{eq: NN delta 0 condit}.  

\begin{figure}[h]
    \centering
    \includegraphics[width=0.3\textwidth]{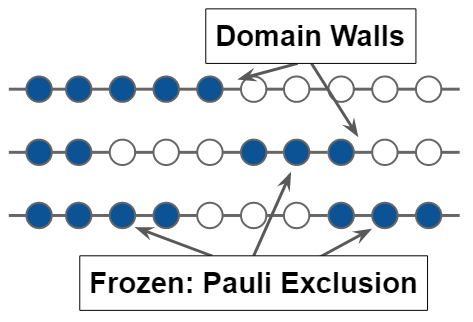}
    \caption{Particle configurations frozen in the Even-Odd NN model at values of $V,\tau$ that satisfy the $\Delta=1$ condition in \eqref{eq: NN delta not 0 condit}.  The only 2-site pairs with a single particle are located on the domain walls.  Since, within the uniform domain, particles are frozen at all times due to Pauli exclusion, the neighboring particle number difference for 2-site pairs on the domain wall is constant and given by $\Delta=1$.  }
    \label{fig:Domain wall freezing}
\end{figure}

However, if every 2-site pair with a single particle is located on the edge of a domain wall in the system, then $\Delta$ will again be well defined (since any neighboring particles will be stationary due to Pauli exclusion) and the conditions \eqref{eq: NN delta not 0 condit} and \eqref{eq: NN delta 0 condit} will hold for these particle configurations.  In Figure \ref{fig:Domain wall freezing}, we give examples of such states that will be stroboscopically frozen when the $\Delta=1$ condition is satisfied, i.e. all these states are eigenstates to the evolution operator ${\cal U}(T) = {\cal T}e^{-i \int_0^T {\cal H}(t)}$.  

We now turn to numerically investigating the emergence of these frozen states and the Hilbert space fragmentation in this system.  We exactly diagonalize ${\cal U}(T)$ at the special points $V=\sqrt{12}$, $\tau=\frac{\pi}{2}$ and $\tau=\pi$ \footnote{As a technical note, the frozen domain wall states will be highly degenerate and numerical diagonalization will give a random basis of eigenstates within the degenerate subspace.  To find the frozen states within this basis, we apply a small disorder potential during the wait step in the evolution to split the energy levels.  This disorder potential will add only a global phase to the frozen states and thus allows a direct numerical route to finding them.}.  Here, the condition for frozen $\Delta=1$ is satisfied, while $\Delta=0$ is perfect swapping or frozen respectively.  If the activated neighboring pairs were disjoint, evolution at these parameter values would be exactly solvable (with dynamics either being a CA or stroboscopically frozen).  As we will see, however, this is not the case here.  The Hilbert space instead fragments into exponentially many subspaces of frozen domain wall states and a single, exponentially large, ergodic subspace.

\begin{figure}[h]
    \centering
    \includegraphics[width=0.495\textwidth]{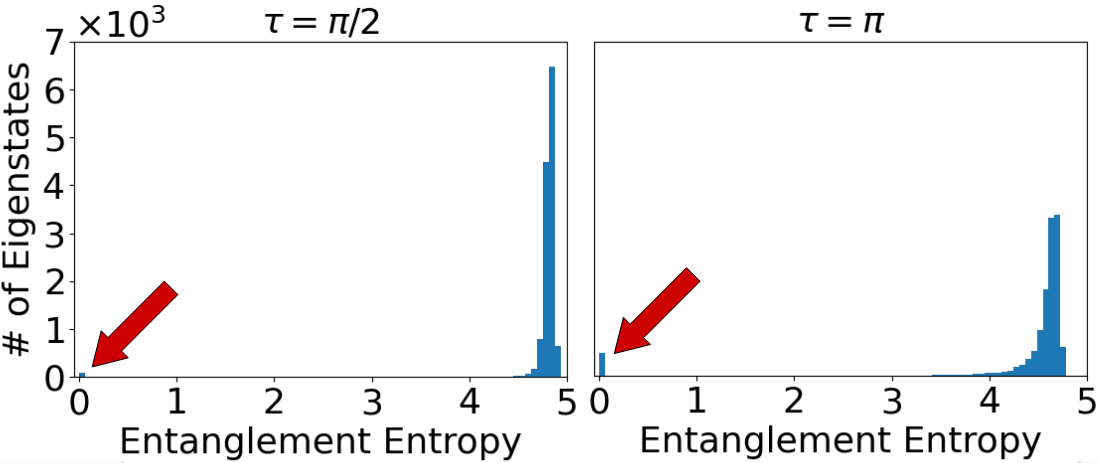}
    \caption{Half-chain entanglement entropy of all eigenstates of the evolution in the even-odd NN Floquet model.   Eigenstates were found by exactly diagonalizing a 16 site chain.  The parameter values were chosen such that both the $\Delta=0$ and $\Delta=1$ conditions in \eqref{eq: NN delta not 0 condit} and \eqref{eq: NN delta 0 condit} are satisfied: $V=\sqrt{12}$,  $\tau=\frac{\pi}{2}$ (left) and $\tau=\pi$ (right).  Despite the non-disjoint nature of the activated hopping site pairs, the conditions \eqref{eq: NN delta not 0 condit} and \eqref{eq: NN delta 0 condit} will still be valid for domain wall states that will, therefore, be frozen under the dynamics.  These number states have no entanglement entropy and are indicated with red arrows in the figure above.  The other eigenstates exhibit near-maximal entanglement entropy.  This is a signature of the fragmentation of the Hilbert space into frozen Krylov subspaces and a ergodic Krylov subspace.}
    \label{fig:domain wall entanglement entropy}
\end{figure}

To seperate the two classes of subspaces, we calculate the half-chain entanglement entropy of the eigenstates (shown in Figure \ref{fig:domain wall entanglement entropy}).  The frozen eigenstates have zero entanglement entropy while the other eigenstates have finite (and as can be seen from Fig. \ref{fig:domain wall entanglement entropy}, large) entanglement entropy.  Upon plotting the average local particle densities of a sample of the zero entanglement entropy eigenstates, we find that they do indeed correspond to the expected frozen domain wall states.  

The large half-chain entanglement entropy of non-domain wall states suggests that the rest of the Hilbert space might be thermalized.  To provide further evidence to this claim, we analyze an indicator often used to differentiate between ergodic and integrable systems: the statistics of level spacing ratios.

For thermalizing systems, it is expected \cite{Dalessio2014Therm} that the evolution operator ${\cal U}$ resembles random matrices drawn from a circular ensemble (the analog of gaussian ensembles for unitary matrices).  Unlike the evolution operators for integrable systems, eigenstates of circular ensembles are random vectors and the spectrum exhibits level repulsion.  Thus, it is possible to argue whether a system is ergodic by analyzing the statistics of the spacing of energy levels to see if the distribution is Poissonian (corresponding to no level repulsion) or if it corresponds to the expected level spacing distribution of circular ensembles (see \cite{Dalessio2014Therm} for explicit formulas).  

Namely, consider the level spacings between two neighboring eigen-quasienergies $\varepsilon$ (i.e. $\varepsilon$'s are the phases of the eigenvalues of ${\cal U}$),

\begin{gather}
    \delta_n = \varepsilon_{n+1} - \varepsilon_n .
    \label{eq: level spacing}
\end{gather}
The ratio of level spacings is given by

\begin{gather}
    r_n = \frac{min\{\delta_n,\delta_{n+1}\}}{max\{\delta_n,\delta_{n+1}\}}.
\end{gather}
We then expect the statistics of $r$ to match that of the circular ensembles instead of yielding a Poissonian distribution if the system is ergodic.  

\begin{figure}[h]
    \centering
    \includegraphics[width=0.495\textwidth]{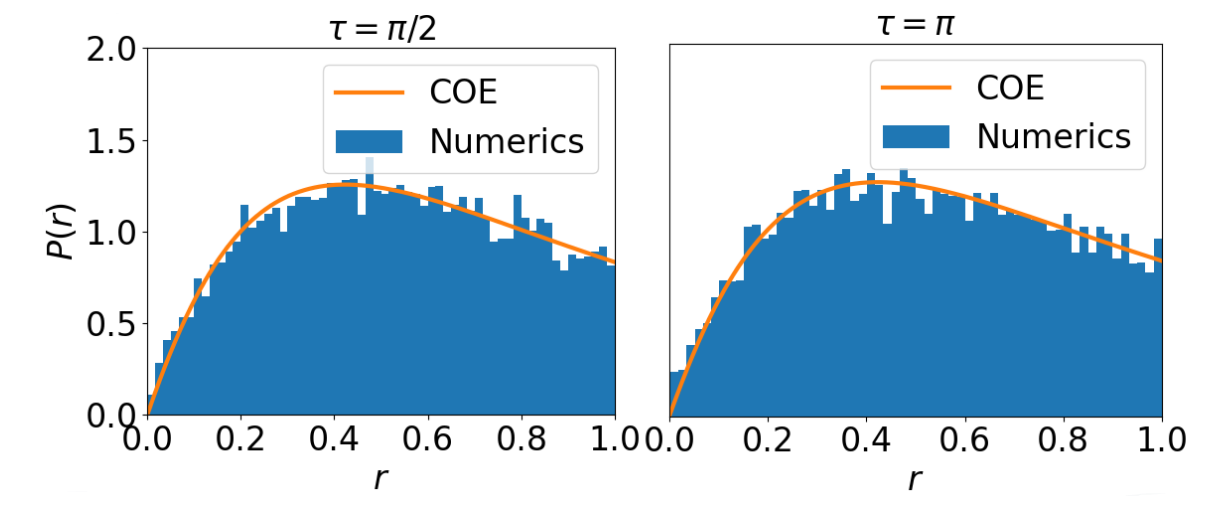}
    \caption{Level spacing statistics in the non-frozen Krylov subspace for evolution in the even-odd NN Floquet model.  As in Fig. \ref{fig:domain wall entanglement entropy}, parameter values are chosen as $V=\sqrt{12}$, $\tau=\frac{\pi}{2}$ (left) and $\tau=\pi$ (right).  The probability distribution, $P(r)$, of the level spacing ratios, $r$, for quasi-energy levels not corresponding to frozen eigenstates provides good agreement with the level spacing probability distribution of random matrices in the circular orthogonal ensemble (COE).  This suggests that the Krylov subspace is ergodic.}
    \label{fig:level spacing}
\end{figure}

In our case, however, the system is not completely ergodic since the domain wall number states are eigenstates of the evolution.  We instead wish to study the nature of the subspace which is the compliment of the set of all frozen Krylov subspaces within the Hilbert space.  We thus will only consider $\delta_n$ in \eqref{eq: level spacing} if the corresponding eigenstates of $\varepsilon_{n+1}$ and $\varepsilon_{n}$ have non-zero half-chain entanglement entropy.  The results of this analysis are shown in Fig. \ref{fig:level spacing}.  As can be seen in the figure, the probability distribution is in good agreement with that of the circular orthogonal ensemble (COE) suggesting that the Krylov subspace is thermal. 

In summary, we have shown that the Hilbert space of the even-odd NN Floquet model is fragmented at special values of interaction strength and driving frequency.  The fragmented Hilbert space simultaneously supports exponentially many (in system size) frozen Krylov subspaces and a single, exponentially large ergodic Krylov subspace.  In this model, we did not find evidence of CA subspaces.  Whether these subspaces are realizable in other non-disjoint models is an open question.  Furthermore, for neighboring two-site pairs each with a single particle, the interactions between the pairs could conspire to produce special values of $V,\tau$ not given by equations \eqref{eq: NN delta not 0 condit} and \eqref{eq: NN delta 0 condit} where evolution is stroboscopically frozen.  We leave both these open questions for future work.

%Looking forward, there are many open questions related to the dynamics and solvability of these non-disjoint, interacting Floquet drives.  Three such questions are (1) is it possible to realize CA Krylov subspaces in such models?  One might imagine scenarios like that presented in Figure \ref{fig: NN 3 bound particles} except with domain wall states evolving into other domain wall states.  (2)      

\section{Summary and discussion}
In recent years the study of quantum many body states that break ergodicity has been an active field of research. Here, we considered conditions for  dynamics in interacting systems that takes initial local number states to local number states. We have found such conditions for systems with sequentially activated hopping involving interactions such as Hubbard and nearest neighbour density interactions. Studying the resultant Diophantine relations between interaction strength, hopping energy, and hopping activation time, we discovered solutions to a variety of such systems. The resultant dynamics can be cast into two types: (1) Evolution that is deterministic for any initial Fock state (2) Fragmentation of the Hilbert space into deterministic sub-spaces and non-deterministic ones.  

Our results introduce new sets of dynamically tractable interacting systems, with an emphasis on 2d where such results are scarce. Furthermore, the approach is applicable to similar systems in other dimensions. At the special solvable points, we get a variaty of behaviors from  frozen dynamics of Fock states to cellular automata like evolution of selected subspaces. 
In cases where only some of the Diophantine conditions are met, we have shown that the special subspaces can exist simultaneously with states that possess volume law entanglement entropy and level statistics suggesting thermalizing behavior. 

As discussed in section \ref{section: Cellular Automation}, although the ratios of Hamiltonian parameters (interaction strength, evolution time etc) considered here are finely tuned, previous work suggests that similarly finely tuned points may be stabilized by disorder to realize novel dynamical phases. In particular, periodic celullar automata evolution in our models may lead to new classes of time crystals.

The problem of finding complete freezing of Fock states also led us to an interesting number theoretic problem involving the solution of a tower of Diophantine equations described in \eqref{eq: NN delta not 0 condit} and \eqref{eq: NN delta 0 condit}. We have shown explicitly solutions for dynamics on lattices with maximal degree of up to $4$ nearest neighbours and conjecture a solution can be found for arbitrary maximal degree. 

We remark that the same methods may be applicable to bosonic systems, and systems with pairing terms where resultant cellular automata may not be of the number preserving type. 

\emph{Acknowledgments.}
We thank G. Refael and E. Berg for discussions. Our work was supported in part by the NSF grant DMR-1918207. 

\bibliography{ArrestedDevelopment.bib}
\bibliographystyle{unsrt}
\onecolumngrid
\appendix

\section{Exact Solutions of Few Site Subspaces}
\subsection{Hubbard Floquet Evolution of 2-site Pair in the 2-particle Sector} \label{appendix: hubbard 2 sector}

We index the 4-particle configurations of the subspace as follows:

\begin{subequations}
\begin{gather}
    0 \rightarrow \hbox{ } \uparrow \downarrow \hbox{\space \space } \text{ \textunderscore \textunderscore}\\
    1 \rightarrow \hbox{ } \text{\textunderscore \textunderscore} \hbox{\space \space \space}\uparrow \downarrow\\
    2 \rightarrow \hbox{ } \uparrow\text{\textunderscore} \hbox{\space\space \space} \text{\textunderscore}\downarrow\\
    3 \rightarrow \hbox{ } \text{\textunderscore}\downarrow \hbox{ } \uparrow\text{\textunderscore}
\end{gather}
\end{subequations}
We therefore have that the representation of the Hubbard Hamiltonian \eqref{eq: RLBL Hubbard Hamiltonian} in this subspace is given by

\begin{gather}
    {\cal H} = \left(\begin{tabular}{c c c c}
         $V$ & $0$ & $-1$ & $-1$ \\
         $0$ & $V$ & $-1$ & $-1$ \\
         $-1$ & $-1$ & $0$ & $0$ \\
         $-1$ & $-1$ & $0$ & $0$ 
    \end{tabular} \right)
\end{gather}
Hence, the evolution, ${\cal U} = e^{-i {\cal H} \tau}$, is given by

\begin{gather}
    {\cal U} = e^{-\frac{1}{2}i V \tau} \left(\begin{tabular}{c c c c}
         $e^{-\frac{1}{2}i V \tau}\left[\frac{1}{2} + A \right]$ & $e^{-\frac{1}{2}i V \tau}\left[-\frac{1}{2} + A \right]$ & $B$ & $B$ \\
         $e^{-\frac{1}{2}i V \tau}\left[-\frac{1}{2} + A \right]$ & $e^{-\frac{1}{2}i V \tau}\left[\frac{1}{2} + A \right]$ & $B$ & $B$ \\
         $B$ & $B$ & $e^{\frac{1}{2}i V \tau}\left[\frac{1}{2} + \Bar{A} \right]$ & $e^{\frac{1}{2}i V \tau}\left[-\frac{1}{2} + \Bar{A} \right]$ \\
         $B$ & $B$ & $e^{\frac{1}{2}i V \tau}\left[-\frac{1}{2} + \Bar{A} \right]$ & $e^{\frac{1}{2}i V \tau}\left[\frac{1}{2} + \Bar{A} \right]$
    \end{tabular} \right)
    \label{eq: U representation for 2 site hubbard}
\end{gather}
where

\begin{gather}
    A(V,\tau) = \frac{e^{\frac{1}{2} i V \tau}}{2} \left[\cos(\frac{1}{2}\tau \sqrt{16+V^2}) - i \frac{V}{ \sqrt{16+V^2}} \sin(\frac{1}{2}\tau \sqrt{16+V^2}) \right] \\
    B(V,\tau) = 2 i \frac{ \sin(\frac{1}{2} \tau \sqrt{16 + V^2})}{\sqrt{16+V^2}}.
\end{gather}
and $\Bar{A}$ is the complex conjugate.

We are now interested in finding when \eqref{eq: U representation for 2 site hubbard} is a permutation matrix.  Note, for non-zero $V$, $|B|<1$.  Thus, our only hope for a permutation matrix is if $B=0$.  This occurs when $\frac{1}{2}\tau \sqrt{16+V^2} = \pi m$ for some $m \in \integers$, i.e. the condition given in \eqref{eq: 2-site permutation condit 1}.

Solving for $A(V,\tau)$ at condition \eqref{eq: 2-site permutation condit 1} yields

\begin{gather}
    A(V,\tau)|_{Condition: \eqref{eq: 2-site permutation condit 1}} = \frac{1}{2} e^{i [\pi m + \frac{1}{2} V \tau]} 
    \label{eq: A after cond 1}
\end{gather}
In \eqref{eq: U representation for 2 site hubbard}, $\cal U$ is a permutation matrix when, in addition to the requirement $B=0$, $|A|=\frac{1}{2}$ and  $\frac{V\tau}{\pi} \in \integers$.  These 2 conditions are uniquely met when, using \eqref{eq: A after cond 1}, $\pi m + \frac{1}{2}V \tau = \pi n$ for some $n \in \integers$.  Thus, we have arrived at the condition given in \eqref{eq: 2-site permutation condit 2}.  When \eqref{eq: 2-site permutation condit 1} and \eqref{eq: 2-site permutation condit 2} are satisfied, $\cal U$ then becomes

\begin{gather}
    {\cal U}|_{\text{Conditions: \eqref{eq: 2-site permutation condit 1} and \eqref{eq: 2-site permutation condit 2}}} = \left(\begin{tabular}{c c c c}
         $n-1$ & $n$ &$0$&$0$\\
         $n$ & $n-1$ &$0$&$0$\\ 
         $0$&$0$&$n-1$ & $n$\\
         $0$&$0$&$n$ & $n-1$\\ 
    \end{tabular}\right)\mod 2
\end{gather}
i.e. yielding the result that when $n$ is even (odd) evolution is the identity (perfect swapping).

\subsection{Nearest Neighbor Floquet Evolution of 2-site Pair in 1-particle Sector}\label{appendix NN RLBL}

The Hamiltonian of the $j^{th}$ 2-site pair is

\begin{gather}
    {\cal H}_j = - ( a_{j1}^\dagger a_{j2} + h.c.) + V n_{j1} n_{j2} + V N_1 n_{j1} + V N_2 n_{j2}
\end{gather}
where $N_1,N_2$ correspond to the number of particles (outside the $j^{th}$ pair) neighboring site $1$ and site $2$ in pair $j$ respectively.  Note, $[N_1,{\cal H}_j] = [N_2,{\cal H}_j] = 0$.  

The representation of the Nearest Neighbor Hamiltonian in the 1-particle sector is given by

\begin{gather}
    {\cal H} = \left(\begin{tabular}{c c}
         $N_1 V$ & $-1$ \\
         $-1$ & $N_2 V$
    \end{tabular} \right)
\end{gather}

Hence, the evolution, ${\cal U} = e^{-i {\cal H} \tau}$, is given by

\begin{gather}
    {\cal U} = \frac{ e^{-\frac{1}{2} i (N_1 + N_2) V \tau}}{C} \left(\begin{tabular}{c c}
         $ C \cos \left[\frac{C}{2} \tau \right] - i \Delta V \sin \left[\frac{C}{2} \tau  \right]$ & $2 i \sin \left[\frac{C}{2} \tau \right]$ \\
         $2 i \sin \left[\frac{C}{2} \tau \right]$ & $ C \cos \left[\frac{C}{2} \tau \right] + i \Delta V \sin \left[\frac{C}{2} \tau \right]$
    \end{tabular} \right)
    \label{eq:U NN}
\end{gather}
where

\begin{gather}
    C(\Delta,V) \equiv \sqrt{4 + \Delta^2 V^2} \\
    \Delta \equiv N_1 - N_2 .
\end{gather}

For perfect swapping to occur, we must have that the diagonal elements of \eqref{eq:U NN} go to zero.  This may only occur when

\begin{gather}
     \frac{C}{2}\tau = \pi (m+\frac{1}{2})\text{ for } m \in \integers \,\text{;}\,  \Delta = 0
     \label{eq: NN swapping condit}
\end{gather}

For freezing to occur, we must have that the off-diagonal elements of \eqref{eq:U NN} are zero.  Note, depending on the particle configuration, $\Delta$ may take any value such that $\Delta \in \integers$ and $|\Delta| < \max \left[\deg( \text{site 1}),\deg(\text{site 2}) \right]$.  We must therefore have that $\frac{C}{2} \tau = \pi m$ for all possible values of $\Delta$ and with $m \in \integers$, i.e. letting $\Delta_i \in \{0,1,...,\max \left[\deg( \text{site 1}),\deg(\text{site 2}) \right]-1 \} $ such that $\Delta_i = \Delta_j$ iff $i = j$, we require

\begin{gather}
    \frac{C(\Delta_i,V)}{2} \tau = \pi m_i \hbox{ } \forall \hbox{ } i
    \label{eq:NN Vt restriction}
\end{gather}
where $m_i \in \integers$.

Combining Equations \eqref{eq: NN swapping condit} and \eqref{eq:NN Vt restriction} yields \eqref{eq: NN delta not 0 condit} and \eqref{eq: NN delta 0 condit} in the main text.

We may now proceed by solving one value of $\Delta_i$ at a time.  We start with $\Delta_0$ and, without loss of generality, let $\Delta_0 \neq 0$ (if $\Delta_0 = 0$, we may simply replace $m_0 \rightarrow \frac{m_0}{2}$ in the final result), we have from \eqref{eq:NN Vt restriction} that 

\begin{gather}
    \tau = \frac{2 \pi m_0}{\sqrt{4 + \Delta_0^2 V^2}}
    \label{eq:first NN restriction}
\end{gather}
Now, looking next at $\Delta_1 \neq 0$ (again, we may set $m_1 \rightarrow \frac{m_1}{2}$ if $\Delta_1 = 0$), we use Eqs. \eqref{eq:NN Vt restriction} and \eqref{eq:first NN restriction} to find

\begin{gather}
    \frac{\sqrt{4 + \Delta_1^2 V^2}}{\sqrt{4 + \Delta_0^2 V^2}} \pi m_0 = \pi m_1 \\
    \implies V = 2 \sqrt{\frac{m_1^2 - m_0^2}{\Delta_1^2 m_0^2 - \Delta_0^2 m_1^2}}
    \label{eq:NN V restriction}
\end{gather}

Now, taking any $\Delta_j$ such that $j\geq 2$ and combining Eqs. \eqref{eq:NN Vt restriction}, \eqref{eq:first NN restriction}, and \eqref{eq:NN V restriction} and simplifying we find 

\begin{gather}
    m_0^2 (\Delta_1^2 - \Delta_j^2) + m_1^2 (\Delta_j^2 - \Delta_0^2) + m_j^2 (\Delta_0^2 - \Delta_1^2) = 0 \text{ ; } m_i \rightarrow \frac{m_i}{2} \text{ if } \Delta_i = 0 
    \label{eq:NN Diophantine appendix}
\end{gather}

Equation \eqref{eq:NN Diophantine appendix} therefore corresponds to a set of $\max \left[\deg( \text{site 1}),\deg(\text{site 2}) \right]-3$ Diophantine equations that must be solved simultaneously to find the values of $m_i$ (and thus $V,\tau$) that correspond to CA dynamics.  Note, also, that in \eqref{eq:NN Diophantine appendix} we must replace $m_i \rightarrow \frac{m_i}{2}$ for whichever $\Delta_i = 0$.      

Note, a particular solution for the first equation in \eqref{eq:NN Diophantine appendix} when $\Delta_0,\Delta_1,\Delta_2 \neq 0$ is $m_0=\Delta_0,m_1=\Delta_1,m_2=\Delta_2$.  Hence, using \eqref{eq: final 3 quadratic dio solution}, the solution to \eqref{eq:NN Diophantine appendix} for $j=2$ is given by

\begin{gather}
    \left(\begin{tabular}{c}
         $m_0$ \\
         $m_1$\\
         $m_2$
    \end{tabular} \right) = d \left(\begin{tabular}{c}
         $-\left[ (\Delta_1^2 - \Delta_2^2) w_1^2 - (\Delta_2^2 - \Delta_0^2) w_2^2 \right] \Delta_0 - 2 (\Delta_2^2 - \Delta_0^2) w_1 w_2 \Delta_1$  \\
         $\left[(\Delta_1^2 - \Delta_2^2) w_1^2 - (\Delta_2^2 - \Delta_0^2) w_2^2 \right]\Delta_1 - 2 (\Delta_1^2 - \Delta_2^2) w_1 w_2 \Delta_0$\\
         $\left[(\Delta_1^2 - \Delta_2^2) w_1^2 + (\Delta_2^2 - \Delta_0^2) w_2^2 \right]\Delta_2$
    \end{tabular} \right)
    \label{eq:NN Dio Solution}
\end{gather}
To obtain the equivalent of \eqref{eq:NN Dio Solution} when, for example, $\Delta_0 = 0$, we must take $m_0 = 2 \Delta_0$ instead of $m_0 = \Delta_0$ in the particular solution of \eqref{eq:NN Diophantine appendix} and relatedly must use $A=\frac{\Delta_1^2-\Delta_2^2}{4}$ instead of $A=\Delta_1^2-\Delta_2^2$ in \eqref{eq: final 3 quadratic dio solution}.  Similar adjustments must be made to $B,m_1$ or $C,m_2$ if $\Delta_1=0$ or $\Delta_2=0$ respectively.  

Equation \eqref{eq:NN Dio Solution} provides all possible solutions for $m_0,m_1,m_2$ (and thus $V,\tau$) that yield classical dynamics for any two site pair with $\Delta = \Delta_0,\Delta_1,$ or $\Delta_2$.  As a corollary, this implies that there exist values of $V,\tau$ (beyond the trivial $V=0$ or $\tau = 0$ solutions) for any measurement protocol that sequentially isolates pairs of sites on a lattice such that all dynamics is a CA so long as the maximum degree of the lattice is at most 3.  In other words, in this case, we may choose $\Delta_0 = 0$, $\Delta_1=1$, and $\Delta_2 = 2$ and (remembering to make the appropriate substitutions since $\Delta_0 = 0$) we find \eqref{eq:NN Dio Solution} becomes \eqref{eq:NN Dio Solution 0 1 2}.  As discussed in the main text, combining \eqref{eq:NN Dio Solution 0 1 2} with the $\Delta=3$ condition yields a new Diophantine equation that may be solved numerically to find non-trivial solutions.  For arbitrary maximal degree, a tower of Diophantine equations emerges.  Whether solutions exist to these Diophantine equations for arbitrary maximal degree (and, if they exist, what they are) we leave as an open problem.   

%We now ask if this is true for lattices of degree at most 4.  For this to be true, we must simultaneously satisfy the $j=2$ and $j=3$ equations \eqref{eq:NN Diophantine appendix}.  To investigate whether this as possible, let us consider the case that $\Delta_0 = 0$, $\Delta_1=1$, and $\Delta_2 = 2$.  By using Eq. \eqref{eq:NN Dio Solution} and remembering to make the appropriate substitution since $\Delta_0 = 0$, we find that
% the $j=2$ equation is satisfied for $m_0,m_1,m_2$ such that  \eqref{eq:NN Dio Solution 0 1 2} holds.  As discussed in the main text, combining Eq. \eqref{eq:NN Dio Solution 0 1 2} and the $\Delta = 3$ condition yields the Diophantine equation \eqref{eq: NN 4 delta Dio} and non-trivial solutions are found numerically. 

\end{document}